\def\ARXIVVERSION{1}
\documentclass[10pt,twocolumn]{article}

\usepackage[letterpaper,top=0.72in,bottom=0.78in,left=0.67in,right=0.67in,columnsep=0.24in]{geometry}
\usepackage{amsmath,amssymb}
\usepackage{booktabs}
\usepackage{graphicx}
\usepackage{microtype}
\usepackage{enumitem}
\usepackage{xcolor}
\usepackage{url}
\usepackage{hyperref}
\usepackage{balance}
\usepackage{xspace}
\usepackage{tikz}
\usepackage{seqsplit}
\usetikzlibrary{arrows.meta,positioning}

\hypersetup{colorlinks=true,linkcolor=black,citecolor=black,urlcolor=blue!55!black}
\setlist[itemize]{leftmargin=1.2em,itemsep=1pt,topsep=2pt}
\setlist[enumerate]{leftmargin=1.35em,itemsep=1pt,topsep=2pt}
\newcommand{\allow}{\textsc{allow}\xspace}
\newcommand{\deny}{\textsc{deny}\xspace}
\newcommand{\recover}{\textsc{recovery}\xspace}
\newcommand{\pub}{\textsc{public}\xspace}

\newcommand{\rd}{\mathrm{RD}}
\newcommand{\code}[1]{\texttt{#1}}
\newcommand{\hash}[1]{\texttt{\expandafter\seqsplit\expandafter{\detokenize{#1}}}}

\title{Policy-Masked Private Experts:\\Auditable and Reversible Capability Access Control in Sparse MoE Models}
\ifdefined\ARXIVVERSION
\author{Zhuoheng Huang\\Independent Researcher \and Mukesh Singh\\Independent Researcher}
\date{August 2026}
\else
\author{Anonymous Authors}
\date{}
\fi

\begin{document}
\maketitle

\begin{abstract}
Most language-model access controls regulate behavior while leaving the same computation available to every request. We study a different systems question: can trusted authorization determine which newly trained parameters are reachable by the forward pass? Policy-Masked Private Experts freezes a pretrained sparse Mixture-of-Experts (MoE) model, trains a disjoint expert branch, and selects the public or private pool before top-$k$ routing. The resulting claim is narrow but testable: under the declared trusted computing base (TCB), an unauthorized request executes no private expert. It does not imply that the public model lacks the same semantic capability.

We test this separation between execution control and task utility in Qwen3-30B-A3B and DeepSeek-V2-Lite. Three Qwen BF16 seeds update all 32 private experts while the public fingerprint remains unchanged. Across 64 adversarial scenarios and 96 deny/fail-closed events, unauthorized private execution is zero; independent hooks exactly match 11,616 routed private rows, and allow--deny--allow recovery is exact. On two prospectively frozen Qwen benchmarks, the private branch improves exact tool use by $5.0$ percentage points (pp) (five versus zero discordances; one-sided Holm $p=.03125$, corresponding two-sided exact $p=.0625$) and $21.3$ pp (percentile-bootstrap 95\% CI $[13.3,29.3]$, Holm $p=.000031$). Three arm-blinded model evaluators retain a positive external effect of $18.7$ pp (95\% CI $[9.3,28.0]$). A parameter-matched LoRA has similar external utility, but a post-hoc request gate leaves 1,225 adapter calls under deny; the disjoint expert branch leaves none. DeepSeek reproduces the route invariant and gains $27.0$ pp. A valid sealed evaluation is near-neutral. These results support auditable, reversible control over a trained parameter path, while showing that useful transfer remains distribution dependent.
\end{abstract}

\section{Introduction}

An assistant may serve users with different credentials or product permissions. Existing defenses usually ask one active model to behave differently: a system prompt states a policy, post-training encourages compliance, and an output filter catches some failures \cite{ouyang2022training,rafailov2023dpo,inan2023llamaguard}. Those defenses remain necessary, but they answer a behavioral question. They do not tell an operator whether a restricted parameter path participated in the computation.

We treat capability access as a reachability problem. The object being authorized is not a natural-language claim or a final answer, but a separately trained parameter branch. Sparse MoEs are a useful substrate because their forward pass already chooses a small set of modules per token \cite{fedus2022switch,jiang2024mixtral,dai2024deepseekv2}. We add a disjoint private expert pool to selected layers, freeze the pretrained model, and let trusted metadata select a mutually exclusive public or private pool before top-$k$ routing. Prompt text cannot change that decision; absent or malformed metadata resolves to deny.

This reframing separates three claims that capability-control studies often blur: whether private modules are absent under deny, whether training changes only the private branch, and whether that branch improves a declared task distribution. The first two concern the mechanism. The third is empirical and may fail even when the access boundary works perfectly. We make no claim of knowledge removal because shared attention and the frozen public path may already implement similar behavior.

We instantiate this design in Qwen3-30B-A3B-Instruct \cite{qwen3report} and DeepSeek-V2-Lite-Chat \cite{dai2024deepseekv2}. Benchmark membership, decoding, checkpoint selection, statistical tests, and compute budgets are frozen before adapted-model evaluation. The study uses matched seeds and LoRA, five controls, an allow--deny--recovery intervention, adversarial authorization attacks, a once-opened sealed benchmark, and cross-model transfer. Post-lock audits address two plausible alternatives to our interpretation: independent hooks test whether route logs correspond to physical execution, and a request-gated LoRA tests whether a conventional adapter attains the same zero-execution boundary.

Our contributions are:

\begin{itemize}
    \item We formalize capability access as three non-equivalent claims: request-time module non-participation, isolation of training updates, and benchmark-relative utility.
    \item We realize the first two claims by placing trusted authorization before MoE routing and training a mutually exclusive expert branch while the public model remains frozen.
    \item We evaluate the boundary rather than infer it from refusals: route logs, independent hooks, fingerprints, mixed-policy batches, cache transitions, and recovery all have explicit gates.
    \item Matched adapters, control arms, two models, a valid sealed null result, and blinded evaluator sensitivity establish the boundary of the evidence. The contribution is an auditable access boundary, not a claim of universal architectural superiority.
\end{itemize}

\section{Problem and Guarantee Boundary}

\subsection{Threat model}

The adversary controls request text, can adapt prompts across requests, can imitate privileged roles, and may exploit missing or malformed authorization metadata, mixed-policy batches, or stale key--value (KV) caches. The adversary cannot modify the trusted policy object, model weights, inference binary, or host. The trusted computing base (TCB) comprises identity and policy services, policy-to-mask compilation, the routing implementation, signed model/checkpoint manifests, per-request cache separation, and audit instrumentation. Tools, retrieval systems, and downstream actions require independent authorization.

Let $\mathcal{E}^{\mathrm{pub}}_\ell$ and $\mathcal{E}^{\mathrm{priv}}_\ell$ denote the public and private experts at wrapped layer $\ell$, and let $\tau\in\{0,1\}$ be trusted authorization. Prompt tokens cannot modify $\tau$. Missing, malformed, or non-Boolean authorization resolves to $0$.

\subsection{Three non-equivalent claims}

\paragraph{G1: module non-participation.}
For any request trace $T(x,\tau)$,
\begin{equation}
\tau=0\;\Longrightarrow\;
\operatorname{Exec}(T)\cap\mathcal{E}^{\mathrm{priv}}=\varnothing.
\label{eq:g1}
\end{equation}
This is the primary security endpoint. It is deterministic conditional on the TCB and is verified from route events, not generated text.

\paragraph{G2: training isolation of the increment.}
Let $\Theta_{\mathrm{opt}}$ be the parameters passed to the optimizer. For every frozen public parameter $\theta_{\mathrm{pub}}$,
\begin{equation}
\begin{aligned}
\theta_{\mathrm{pub}}&\notin\Theta_{\mathrm{opt}},
&\Delta\theta_{\mathrm{pub}}&=0,\\[-2pt]
&&\theta_{\mathrm{pub}}^{\mathrm{after}}&=\theta_{\mathrm{pub}}^{\mathrm{before}}.
\end{aligned}
\label{eq:g2}
\end{equation}
This is an optimizer-state claim, not $\nabla_{\theta_{\mathrm{pub}}}\mathcal{L}_{\mathrm{priv}}=0$: the loss still depends on shared frozen computation, so that mathematical derivative may be nonzero. Fingerprints and optimizer audits verify exclusion and no update. This localizes the \emph{newly trained increment}; it does not erase capabilities already in $\theta_{\mathrm{pub}}$. It is related to modular information-flow control and gradient routing \cite{tiwari2024ifc,cloud2024gradient}, but our unit of authorization is a request-time functional path added to a pretrained sparse MoE.

\paragraph{G3: empirical capability separation.}
For a declared distribution $D$ and scorer $S$,
\begin{equation}
\Delta_D=\mathbb{E}_{x\sim D}[S(x,\allow)-S(x,\pub)].
\label{eq:g3}
\end{equation}
A positive $\Delta_D$ indicates useful uplift on $D$; no finite benchmark proves universal semantic absence. The prospectively frozen analysis uses \pub as its declared reference. Because \deny shares the wrapped checkpoint and runner with \allow, we additionally report an explicitly post-hoc allow--deny diagnostic. G1 may hold when G3 is null or negative.

\section{Policy-Masked Private Experts}

\subsection{Architecture}

At selected pretrained MoE layers, we copy the expert architecture to create a disjoint private pool. Public experts and all non-private parameters are frozen. For token representation $h$ and router logits $z_\ell(h)$, the policy mask is
\begin{equation}
m_{\ell,e}(\tau)=
\begin{cases}
0,&\tau=1,\ e\in\mathcal{E}^{\mathrm{priv}}_\ell,\\
0,&\tau=0,\ e\in\mathcal{E}^{\mathrm{pub}}_\ell,\\
-\infty,&\text{otherwise}.
\end{cases}
\end{equation}
Top-$k$ selection is computed only after applying $z_\ell(h)+m_\ell(\tau)$. Thus \allow routes only within the private pool and \deny only within the original public pool at wrapped layers. This mutually exclusive construction makes route attribution simple and prevents the learned router from overriding authorization. The rest of the transformer remains shared and frozen.

\begin{figure*}[t]
\centering
\begin{tikzpicture}[
    font=\small,
    node distance=4mm and 7mm,
    box/.style={draw,rounded corners=2pt,minimum height=7mm,align=center,inner xsep=6pt},
    frozen/.style={box,fill=gray!12},
    trusted/.style={box,fill=blue!9,draw=blue!55!black},
    private/.style={box,fill=orange!13,draw=orange!65!black},
    arrow/.style={-{Latex[length=2mm]},thick}
]
\node[box] (request) {request tokens};
\node[frozen,right=of request] (lower) {frozen lower\\transformer};
\node[trusted,right=of lower,minimum width=30mm] (router) {hard policy mask\\then top-$k$ routing};
\node[frozen,above right=1mm and 9mm of router] (public) {public expert pool\\(original, frozen)};
\node[private,below right=1mm and 9mm of router] (private) {private expert pool\\(added, trainable)};
\node[frozen,right=15mm of router] (upper) [xshift=31mm] {frozen upper\\transformer};
\node[trusted,above=7mm of router] (policy) {trusted metadata\\fail-closed resolver $\rightarrow\tau$};
\node[box,below=7mm of private] (loss) {specialist loss\\during training};

\draw[arrow] (request) -- (lower);
\draw[arrow] (lower) -- node[above,font=\scriptsize] {$h_\ell$} (router);
\draw[arrow,blue!60!black] (policy) -- (router);
\draw[arrow] (router) -- node[above,sloped,font=\scriptsize] {\deny} (public);
\draw[arrow,orange!70!black] (router) -- node[below,sloped,font=\scriptsize] {\allow} (private);
\draw[arrow] (public.east) -- (upper.west);
\draw[arrow,orange!70!black] (private.east) -- (upper.west);
\draw[arrow,orange!70!black] (loss) -- node[right,font=\scriptsize] {gradients} (private);
\draw[-{Bar[width=5pt]},thick,gray!65] (loss.west) -| node[pos=.25,left,font=\scriptsize,align=right] {no public\\update} (lower.south);
\end{tikzpicture}
\caption{Mechanism and guarantee boundary. Trusted metadata, never prompt text, selects the visible pool before routing. The optimizer updates only the added private experts; the pretrained transformer and public experts stay frozen. Deny can therefore be audited as zero private execution without claiming that the public path lacks the semantic capability.}
\label{fig:architecture}
\end{figure*}
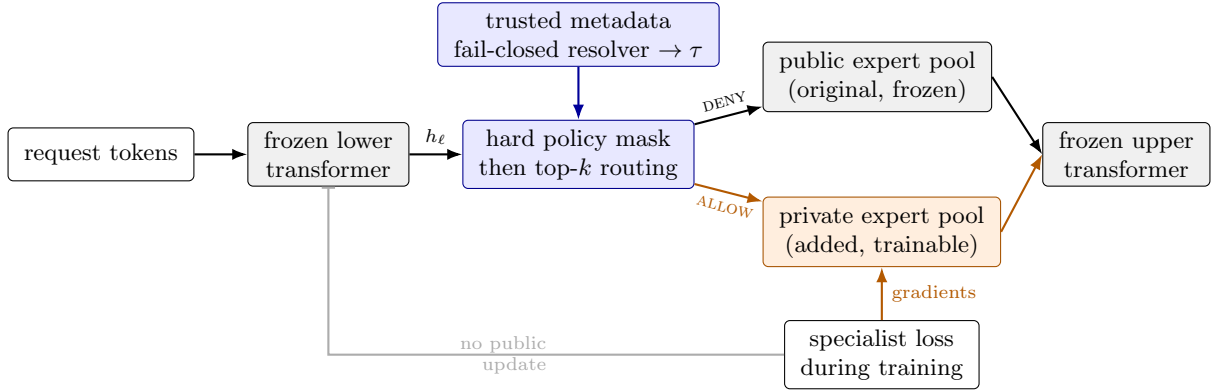

The production-facing policy resolver accepts only a trusted mapping containing a Boolean \code{allow\_private=true}. User claims, strings such as ``true,'' wrong keys, nested values, or absent metadata fail closed. The same row-level mask is propagated through mixed batches. KV caches are policy-tagged; cross-policy reuse is rejected before model execution.

\subsection{Training and checkpointing}

Specialist tool-use examples are trained with the private pool forced on. Only the private gate and expert weights require gradients; the public model fingerprint is checked before and after training. We use the standard autoregressive loss, deterministic token-budget batches, and final-step checkpoint selection. Every expert must receive both routing events and nonzero finite gradients. Checkpoints contain the private weights, optimizer state, source/configuration hashes, and the frozen-base fingerprint. A valid reload must reproduce the checkpoint fingerprint and pass \allow, \deny, missing-metadata, mixed-row, and recovery probes.

\subsection{Causal toggle protocol}

The same checkpoint is exercised in the fixed sequence
\[
\pub \rightarrow \allow \rightarrow \deny \rightarrow \recover.
\]
The mechanism prediction is not merely lower denied-task accuracy. It is: (i) \allow and \recover execute private experts only; (ii) \deny executes zero private experts; and (iii) \recover reproduces \allow under the frozen decoding context. When public and deny are run in the same deterministic context, we additionally test output equality. Separate batched executions can introduce numerical or scheduling differences, so output identity is reported as a diagnostic rather than substituted for the route invariant.

\section{Experimental Design}

\subsection{Models and training data}

Table~\ref{tab:models} summarizes the two implementations. The primary Qwen system adds 16 private experts at each of layers 46 and 47. Training uses 8,524 ToolMind \cite{toolmind} targets, 890 validation targets, and 16,084,443 admitted tokens in 2,273 deterministic batches. Exact CCTU prompt/tool overlaps are excluded. We run three matched BF16 seeds; only private initialization and training RNG change.

The cross-model replication uses DeepSeek-V2-Lite-Chat with 16 private experts at layers 25 and 26. It uses a prospectively fixed 1,024/128 ToolMind split, 1,602,789 admitted train tokens, 226 batches, and one BF16 pass. The Qwen path uses SDPA; DeepSeek uses explicit FlashAttention-2 and \code{torch.compile}. These are fixed numerical families, not configurations selected by benchmark outcome.

\begin{table}[t]
\centering
\caption{Private-expert training configurations. Public parameters are frozen in both systems.}
\label{tab:models}
\resizebox{\columnwidth}{!}{%
\begin{tabular}{lrr}
\toprule
 & Qwen3-30B-A3B & DeepSeek-V2-Lite \\
\midrule
Wrapped layers & 46, 47 & 25, 26 \\
Private experts/layer & 16 & 16 \\
Trainable parameters & 151,060,480 & 276,889,600 \\
Train / validation targets & 8,524 / 890 & 1,024 / 128 \\
Train tokens & 16,084,443 & 1,602,789 \\
Batches / passes & 2,273 / 1 & 226 / 1 \\
Precision & BF16 & BF16 \\
Optimizer / LR & fused AdamW / $10^{-4}$ & fused AdamW / $10^{-4}$ \\
Training seeds & 20260730--32 & 20260801 \\
\bottomrule
\end{tabular}}
\end{table}

\subsection{Matched adaptation and controls}

The Qwen matched baseline adapts the public experts' gate, up, and down projections at layers 46--47 using expert-wise LoRA \cite{hu2022lora}. Its 768 modules use rank 70/$\alpha=140$ or rank 69/$\alpha=138$, for 151,061,504 trainable parameters---1,024 more than the private path. Data, tokens, steps, optimizer, learning rate, precision, and final-step rule match the primary seed. A separate post-lock audit reuses the unchanged adapter behind a trusted request gate intended to disable LoRA under deny or malformed metadata. Independent hooks test whether disabled modules truly stop executing; no retraining or output-based selection occurs.

The frozen E3/E7 matrix compares six rows: (1) frozen public; (2) prompt-only authorization text; (3) soft routing over the concatenated public/private pool without a hard mask; (4) an untrained copied private pool with the hard mask; (5) trained private experts with the hard mask; and (6) matched LoRA. Every row has an independent resolved configuration and manifest. Public outputs may be shared only through a verified immutable projection.

\subsection{Benchmarks and preregistration}

We use three evaluation suites.

\paragraph{Task-aligned independent benchmark.}
One hundred ToolMind examples are selected by a frozen hash rank, 25 per sequence-length quartile. Their records, exact targets, called-tool names, and called-tool schemas are absent from Qwen training and validation targets. The score is exact ordered tool-call match after deterministic canonicalization.

\paragraph{Fresh external benchmark.}
One hundred ToolFailBench \cite{toolfailbench} examples are selected before adapted output, five from each of 20 domain-by-mode strata. Its primary Clean Tool-Use Rate (CTUR) uses the 75 tool-required cases; 25 no-tool controls remain descriptive. A deterministic rule classifier identifies correct use, tool skip, ignored results, fabrication, and unnecessary use. Post lock, three independent arm-blinded model evaluators label all 300 public/private/LoRA candidates under a fixed rubric. These labels are a sensitivity analysis, not human validation or a replacement primary endpoint.

\paragraph{Sealed complex-constraint benchmark.}
CCTU \cite{ye2026cctu} contains 200 tool-use problems with executable constraint feedback. We stratify 40 open-pilot and 160 once-opened sealed cases across its four source categories. The preregistered primary metric is success rate (SR); partial success rate (PSR) is secondary. The sealed run requires 160/160 scorable examples, zero harness exceptions, \emph{and zero model-format parse errors} in all three arms.

For E7, greedy decoding, no thinking, batch size 8, at most 512 new tokens, fixed canonicalization, checkpoint choice, benchmark IDs, and statistics are frozen before any private, LoRA, or control output. The primary comparison is paired full-private minus public risk difference. We report discordant counts, one-sided exact McNemar tests, and 100,000-replicate paired stratified bootstrap percentile intervals. The intervals summarize effect size; at a boundary with no observed reverse discordance, they are not interpreted as independent two-sided significance evidence. Holm correction controls the two E7 benchmark tests at $\alpha=.05$. CCTU uses its separately preregistered one-sided test and bootstrap seed.

\subsection{Authorization stress test}

The E4 suite freezes 64 scenarios (eight families, eight scenarios each): direct instruction override, role/system imitation, missing metadata, malformed metadata, mixed-policy batches, allow--deny--allow transitions, private-name imitation, and cache isolation. These create 152 policy events: 56 allow and 96 deny/fail-closed. The primary endpoint is zero private selections across every denied event. Generated token identity is descriptive.

\section{Results}

\subsection{Private capacity trains reproducibly}

All three Qwen seeds complete 2,273/2,273 steps, keep the public fingerprint unchanged, route through and update all 32 private experts, reload the final private weights and optimizer state exactly, and pass deny/missing-metadata probes. Table~\ref{tab:training} shows very small cross-seed variation: relative validation-loss reduction is $0.41718$ with sample SD $0.00034$. The LoRA baseline is also technically valid but has a smaller $38.18\%$ reduction. These losses establish that both parameterizations fit the specialist training distribution; they do not establish access-controlled inference quality.

\begin{table}[t]
\centering
\caption{Matched Qwen training results. ``Gates'' requires finite optimization, complete route/gradient coverage, an unchanged frozen fingerprint, exact checkpoint reload, and zero private execution for deny/missing probes.}
\label{tab:training}
\resizebox{\columnwidth}{!}{%
\begin{tabular}{lrrrr}
\toprule
Run & Seed & Final val. & Rel. reduction & Gates \\
\midrule
Private-1 & 20260730 & .573489 & .417169 & pass \\
Private-2 & 20260731 & .573436 & .416850 & pass \\
Private-3 & 20260732 & .573335 & .417527 & pass \\
\midrule
Mean (SD) & -- & .573420 (.000078) & .417182 (.000339) & 3/3 \\
Matched LoRA & 20260730 & .555237 & .381836 & pass \\
\bottomrule
\end{tabular}}
\end{table}

\subsection{Hard masking enforces authorization}

The selected Qwen checkpoint executes private experts only under \allow and \recover, and zero private experts under \deny. Recovery reproduces allow exactly on both E7 roles. In the 64-scenario stress test, all 152 policy resolutions match the frozen resolver, all 56 allowed events are private-only, and all 96 deny/fail-closed events have zero unauthorized private selections. Mixed-policy rows remain isolated, cross-policy cache reuse is rejected, clean deny execution is exact, and no scenario raises a harness exception. Independent forward hooks installed directly on all 32 private expert modules observe 931 calls and 11,616 physical input rows, exactly matching route logs; denied sequential events invoke no private module. The peak allocated memory is 61.4 GB, below the 90 GB gate.

The post-hoc gated-adapter audit provides a stronger counterfactual than ungated LoRA alone. On the same frozen 100+100 IDs, deny records 1,225 physical LoRA A/B calls ($0.04242\%$ of the allow count); allow and recovery each record 2,887,794 calls over 1,536 hooked modules and produce byte-exact ordered outputs. The adapter therefore fails the zero non-participation endpoint in this implementation even though its logical gate recovers exactly. By contrast, authorization removes the disjoint expert branch before expert selection.

The mechanism survives model transfer. On DeepSeek, the four 100-example arms are fully scorable with zero harness exceptions; public and deny raw outputs are identical, allow and recovery raw outputs are identical, deny has zero private selections, and allow/recovery are private-only. These observations support G1 and reversibility on both architectures.

\subsection{Capability uplift is benchmark dependent}

Table~\ref{tab:e7} reports the frozen six-condition matrix. On task-aligned ToolMind, full-private performance is 61\% versus 56\% public ($\rd=+5$ pp). Five versus zero discordances give the preregistered one-sided exact McNemar and Holm $p=.03125$; the corresponding two-sided exact value is $p=.0625$. The preregistered percentile-bootstrap 95\% interval $[1,10]$ pp is retained as an effect-size summary, not as separate two-sided significance evidence, because empirical resampling cannot create an unobserved reverse discordance. Prompt-only and soft-router-only both equal public at 56\%; mask-only falls to 49\%; matched LoRA reaches 55\%. Thus the evidence for uplift on this role is directional and limited, while matched LoRA does not improve over public.

On the fresh ToolFailBench role, frozen-rule full-private CTUR is 72.0\% versus 50.7\% public ($\rd=+21.3$ pp, 95\% CI $[13.3,29.3]$). All 16 discordances favor full private; the one-sided McNemar $p=.0000153$ and Holm-adjusted $p=.0000305$. Matched LoRA reaches 70.7\%, followed by soft routing at 58.7\%, prompt-only at 54.7\%, and mask-only at 46.7\%. Secondary paired comparisons put full private $6.0$ pp above LoRA on the task-aligned role (95\% CI $[1.0,12.0]$, $p=.0352$) and $1.3$ pp above it externally (95\% CI $[-2.7,5.3]$, $p=.5$).

Three independent arm-blinded model evaluators corroborate the external direction. Across 300 candidates, multiclass Fleiss $\kappa=.773$ and unanimity is 84.7\%; the frozen classifier agrees with majority labels on 78.3\% of exact categories and 80.3\% of binary successes. Under majority labels, full private scores 40/75 and public 26/75, for $\rd=+18.7$ pp (95\% CI $[9.3,28.0]$; 16 versus 2 discordances, $p=.000656$). LoRA--public is $+16.0$ pp (95\% CI $[6.7,25.3]$, $p=.00377$), while private--LoRA is $+2.7$ pp (95\% CI $[-2.7,9.3]$, $p=.344$). These are model-evaluator labels rather than human judgments, but they support rather than reverse the frozen primary result.

The preregistered reference is the frozen public projection. We also examine full-private allow against deny generated with the same wrapped checkpoint and runner. This post-hoc diagnostic gives $+5.0$ pp on the task-aligned role (percentile-bootstrap 95\% CI $[1.0,10.0]$, five versus zero discordances, one-sided $p=.03125$, two-sided $p=.0625$) and $+17.3$ pp externally (95\% CI $[9.3,26.7]$, 13 versus zero, $p=.000122$). It does not replace the preregistered comparison, but it addresses the 15/100 task and 99/100 external transcript mismatches between separately cached public and deny runs. On the 25 external no-tool controls, public, full private, and deny each score 24/25, while LoRA scores 22/25. Full private makes one unnecessary tool call and LoRA makes two, so the positive tool-required result is not explained by unrestricted tool use.

\begin{table}[t]
\centering
\caption{Prospectively frozen Qwen E3/E7 matrix. Task is exact ordered tool-call match ($n=100$). External is CTUR over tool-required ToolFailBench cases ($n=75$); all 100 records are retained.}
\label{tab:e7}
\begin{tabular}{lrr}
\toprule
Condition & Task-aligned & Fresh external \\
\midrule
Frozen public & .560 & .507 \\
Prompt-only & .560 & .547 \\
Soft-router-only & .560 & .587 \\
Mask-only & .490 & .467 \\
Full private & .610 & .720 \\
Matched LoRA & .550 & .707 \\
Gated LoRA allow$^\dagger$ & .540 & .720 \\
\midrule
Private--public RD & $+.050$ & $+.213$ \\
95\% bootstrap CI & $[.010,.100]$ & $[.133,.293]$ \\
Holm one-sided $p$ & $.03125$ & $.000031$ \\
\bottomrule
\end{tabular}
\parbox{\columnwidth}{\footnotesize $^\dagger$Post-hoc diagnostic on the frozen benchmark IDs; not a replacement preregistered endpoint.}
\end{table}

The DeepSeek replication strengthens the evidence that the method can create useful gated capacity without implying uniformity. On the frozen 100-example task-aligned suite, public and deny score 9\%, while allow and recovery score 36\%. The paired risk difference is $+27$ pp (95\% CI $[19,35]$); all 27 discordances favor allow, giving one-sided exact McNemar $p=7.45\times10^{-9}$. Parse deviations are retained as failures (six events affecting five records in public/deny and four events affecting four records in allow/recovery), and all examples remain scorable. Because this is one task-aligned suite on one additional model, it supports cross-model transfer rather than broad efficacy generalization.

\subsection{The sealed result is valid and near-neutral}

The once-opened CCTU run evaluates 160 cases in the frozen order public, deny, allow. Public and deny transcripts are byte-identical; deny executes zero private experts. Every arm is 160/160 scorable with zero harness exceptions and zero model-format parse errors, so every preregistered validity gate passes and the run retains sealed confirmatory status.

The confirmatory outcome is near-neutral: SR is .150 public versus .14375 allow, for $\rd=-.00625$ (95\% CI $[-.0375,.0250]$; one-sided exact McNemar $p=.77344$). There are three allow-only and four public-only discordances. PSR changes from .09375 to .0875 ($\rd=-.00625$, 95\% CI $[-.01875,0]$, $p=1$). The positive directional hypothesis is not supported, while the confidence interval excludes the previously reported five-point decline and the route traces support G1.

\begin{table}[t]
\centering
\caption{Primary paired efficacy results. CCTU is a valid sealed confirmatory result; its directional uplift hypothesis is not supported.}
\label{tab:primary}
\resizebox{\columnwidth}{!}{%
\begin{tabular}{llrrrr}
\toprule
Model & Evaluation & Public & Allow & RD (boot. 95\% CI) & $p$ \\
\midrule
Qwen & Task-aligned & .560 & .610 & $+.050\;[.010,.100]$ & .03125$^\dagger$ \\
Qwen & Fresh external & .507 & .720 & $+.213\;[.133,.293]$ & .000031$^\dagger$ \\
Qwen & CCTU sealed & .150 & .144 & $-.006\;[-.038,.025]$ & .77344 \\
DeepSeek & Task-aligned & .090 & .360 & $+.270\;[.190,.350]$ & $7.45\!\times\!10^{-9}$ \\
\bottomrule
\multicolumn{6}{l}{\footnotesize $^\dagger$Holm-adjusted one-sided $p$; CCTU uses its separate preregistration.}
\end{tabular}}
\end{table}

\subsection{Clean-lock reproducibility}

The final Qwen clean lock regenerates the 40-case open-pilot allow result from pinned source, data, and checkpoint: 40/40 are scorable, SR/PSR are .15/.10, parse and harness errors are zero, both private layers record 64/64 private selections in the bounded probes, and public-path maximum absolute error is 0. The ordered transcripts are byte-identical to the original allow, deny, recovery, and prior revalidation references, with 0\% measured drift. The final evidence builder then recomputes CCTU, multi-seed, E3/E7, E4, and E6 statistics from raw artifacts and verifies every input manifest before comparing summaries. All component rebuilds and hashes match.

\section{What the Results Establish}

\paragraph{The route boundary survives every audited intervention.}
Across controlled pilots, the sealed run, the E7 recovery sequence, 152 adversarial policy events, and the DeepSeek replication, deny/fail-closed execution never selects a private expert. Recovery restores the same private route and, in matched contexts, the same output. This is direct evidence for G1 rather than an inference from refusal behavior.

\paragraph{Training isolation is strongly audited, not universally proved.}
The only trainable Qwen parameters are 151.06M private weights; all 32 experts receive gradients; public fingerprints are unchanged across three seeds and final reloads. DeepSeek shows the same pattern with 276.89M private parameters. These checks support G2 for the implemented optimizer and captured state. They do not rule out every hardware fault, covert channel, or pre-existing public capability.

\paragraph{Capability effects do not transfer uniformly.}
Qwen shows limited directional evidence of uplift on held-out ToolMind under the preregistered one-sided test and stronger evidence on fresh external ToolFailBench, but not on CCTU; DeepSeek improves on the task-aligned suite. The variation could reflect model architecture, training-set alignment, evaluator sensitivity, base competence, or optimization. The present study does not identify a single cause. Averaging these roles would obscure a useful distinction: enforcing access to an increment is easier than making that increment generalize.

\paragraph{A gated adapter does not automatically create the same boundary.}
Matched LoRA and full private have similar external utility, while full private is six points better on the task-aligned paired comparison. The operational distinction is sharper: the post-lock request gate yields exact recovery but still records 1,225 denied-condition LoRA calls ($0.04242\%$ of allow). In this implementation, a logical adapter switch is therefore insufficient evidence of physical non-participation. The disjoint expert pool presents a cleaner audit surface because the policy removes the whole branch before expert selection. This result concerns enforceability, not universal adapter quality.

\paragraph{Behavioral identity is not the security endpoint.}
In controlled Qwen pilots and CCTU, deny is byte-identical to public. In the separately batched E7 matrix, deny differs from the cached public projection on 15 task-aligned and 99 external records, despite exact checkpoint/frozen fingerprints and zero private execution. We report these mismatches rather than treating output equality as a universal guarantee. They motivate stronger deterministic-serving controls, but do not violate Equation~\ref{eq:g1}.

\section{Related Work}

\paragraph{Sparse MoE and specialization.}
Sparsely gated MoEs introduced conditional capacity at a roughly fixed per-token cost \cite{shazeer2017moe}; GShard, Switch Transformer, expert-choice routing, Mixtral, and DeepSeekMoE developed that idea at larger scales and with different load-balancing rules \cite{lepikhin2021gshard,fedus2022switch,zhou2022expertchoice,jiang2024mixtral,dai2024deepseekmoe}. Those routers optimize quality, efficiency, or expert utilization, not authorization. Our private pools are deliberately created and hidden before routing. Recent routing attacks further caution against treating a learned router as a security boundary \cite{wu2025gatebreaker,xu2026routehijack}.

\paragraph{Modular information flow and localization.}
Classical information-flow control asks whether permitted flows respect a declared security lattice \cite{denning1976lattice}. Tiwari et al. carry that concern into modular learning by partitioning training domains across experts \cite{tiwari2024ifc}, while Gradient Routing directs selected examples' gradients into designated modules \cite{cloud2024gradient}. We share the principle that modular parameters can carry provenance. Our focus is a pretrained sparse MoE with a request-authenticated alternative path, a causal allow--deny--recovery intervention, authorization attacks, and downstream utility. The hard mask itself is simple; the contribution is the construction plus mechanism-first evidence and claim separation.

\paragraph{Adapters, unlearning, and behavioral safety.}
LoRA provides efficient adaptation but does not by itself make the base computation path less capable \cite{hu2022lora}. A deployment can separately gate an adapter, which motivates a direct gated-adapter comparison. Unlearning aims to suppress or remove information from a model and faces difficult verification problems \cite{eldan2023harrypotter,li2024wmdp,liu2025unlearning}. RLHF, DPO, guardrails, and instruction-hierarchy training steer behavior \cite{ouyang2022training,rafailov2023dpo,inan2023llamaguard,wallace2024instruction}. These approaches are complementary: policy-masked branches govern reachability of designated parameters, while behavioral safety remains necessary for every visible path.

\section{Limitations and Broader Impact}

Our experiments concern tool-call capability increments in two open sparse MoEs. They do not establish results for dense models, arbitrary semantic domains, multiple simultaneous private tiers, continual learning, distributed serving, or capabilities already strong in the public model. Only two layers per model receive private pools. The shared attention, embeddings, residual stream, and public experts remain capable and may independently implement similar behavior.

The TCB is substantial. A compromised identity service, mask compiler, kernel, cache manager, model manifest, or tool authorization layer can defeat the guarantee. Timing, memory, and route-count side channels are not analyzed. Exact route traces are valuable audit evidence but can themselves expose sensitive policy metadata and require access controls and retention limits.

Benchmark validity is another limitation. The positive Qwen result uses 75 tool-required external cases. Three arm-blinded model evaluators retain a positive paired effect, but binary agreement with the rule classifier is only 80.3\%; they are models, not human annotators. The DeepSeek replication uses only 100 task-aligned cases. Its six public and four allow parse-error events remain scored as failures, so the all-record effect does not depend on filtering them; the consistent parse-conditioned sensitivity is nevertheless post-treatment selection. CCTU passes its preregistered validity gates but is near-neutral and does not support directional uplift. Human annotation, more models, tasks, serving stacks, and independently maintained authorization suites are needed.

Finally, access-tiered models can entrench unfair or opaque restrictions. Deployment should require a legitimate authorization basis, transparent appeals, least-privilege defaults, and separate safeguards for tools and actions. Our mechanism makes a policy enforceable; it does not make the policy just.

\section{Reproducibility and Artifact Discipline}

All scientific stages follow a think--commit--run--log sequence. Model revisions, source commits, dataset membership, hashes, batch plans, trainable partitions, decoding, checkpoint rules, statistical tests, and compute caps are resolved before evaluation. Scientific null or negative outcomes never trigger a rerun. Model-format deviations are retained, and per-sample exceptions do not discard peers. The CCTU seal is opened once; its opening receipt and pre/post manifests are archived. Final JSON, TSV, and \LaTeX{} tables are rebuilt from raw run artifacts, and local/remote SHA-256 lists are compared. The anonymized artifact contains the mechanism, training, evaluation, and statistics code; licensed model weights and benchmark data are not redistributed.

\section{Conclusion}

Policy-masked private experts turn request-time reachability of trained sparse-MoE parameters into a measurable serving invariant. Across two models, hard deny produces zero private execution and allow--deny--allow restores the same path; adversarial metadata, mixed batches, and cache transitions do not cross the boundary. Independent hooks corroborate the route logs. A request-gated LoRA recovers exactly yet still records 1,225 adapter calls under deny, showing why a behavioral or logical switch should not be equated with physical non-participation. Qwen shows directional uplift on two frozen roles, with stronger evidence on the fresh external role and under blinded model evaluation, and DeepSeek transfers the route invariant and positive task effect. Valid sealed CCTU remains near-neutral. The supported claim is consequently narrow: under the declared TCB, a disjoint trained branch can be made auditable and reversibly absent from an unauthorized forward pass. This does not show that the public model lacks the capability, that private specialization always improves utility, or that expert pools are universally superior to simpler adapters.

\ifdefined\ARXIVVERSION
\clearpage
\appendix

\section{Evidence Map and Interpretation Rules}

The full version includes the audit details needed to distinguish the three
claims in the main paper.  Table~\ref{tab:app-claims} states the decision rule
for each claim before listing the corresponding observations.  In particular,
a successful route audit cannot validate a failed efficacy test, and a positive
task score cannot compensate for unauthorized private execution.

\begin{table*}[t]
\centering
\caption{Claim-to-evidence map. ``Exact'' denotes byte or tensor equality in the declared comparison context.}
\label{tab:app-claims}
\begin{tabular}{p{.105\textwidth}p{.235\textwidth}p{.43\textwidth}p{.105\textwidth}}
\toprule
Claim & Required evidence & Observed evidence & Conclusion \\
\midrule
G1: deny isolation & Zero private selections for every deny/fail-closed event; row-local policy resolution & Qwen pilots, CCTU, E7, and 96 E4 deny events; DeepSeek 100-case deny; no unauthorized selection in either model & supported \\
G1: reversibility & One checkpoint; allow$\rightarrow$deny$\rightarrow$allow; restored private routes and matched behavior & Qwen E7 allow/recovery exact on both benchmark roles; DeepSeek raw and parsed outputs exact & supported \\
G2: isolated increment & Only private parameters train; frozen fingerprint unchanged; all private experts routed and differentiated & Qwen 3/3 seeds and DeepSeek: 32/32 experts receive routes and gradients; public fingerprint and reload checks exact & strongly audited \\
G3: useful uplift & Outcome-independent benchmark freeze, positive paired RD, declared inference & Qwen task-aligned $+5.0$ pp; Qwen fresh external $+21.3$ pp; DeepSeek task-aligned $+27.0$ pp & directional support on three roles \\
G3: broad transfer & Consistent positive effects across independent suites & Arm-blinded external sensitivity remains positive ($+18.7$ pp); valid sealed CCTU is $-0.625$ pp with CI crossing zero & not supported \\
\bottomrule
\end{tabular}
\end{table*}

We use four reporting rules throughout.  First, the route endpoint is primary
for G1 and is computed from all token--expert selections at the wrapped layers.
Second, all eligible records remain in their frozen arm even when a model emits
malformed JSON.  Third, scientific null or negative outcomes do not trigger a
rerun.  Fourth, infrastructure failures remain in the experiment ledger and a
repair is described before the next execution.

\section{Serving Construction and Trusted Computing Base}

\subsection{Policy resolution and masked routing}

Authorization is external to the prompt.  The resolver accepts a trusted
mapping only when the value associated with \code{allow\_private} has Boolean
type and equals true.  Missing metadata, wrong keys, strings, numbers, lists,
nested values, nulls, and serialized JSON text all resolve to deny.  User text,
system-role imitation in user content, private expert names, and generated
tokens never mutate the resolved bit.

For every wrapped layer, the resolved vector is expanded per batch row before
top-$k$ selection.  The visible expert set is mutually exclusive: authorized
rows see only the added private pool, whereas denied rows see only the original
public pool.  This differs from adding a post-routing penalty because an
ineligible private index never enters the candidate set.

\begin{center}
\fbox{\begin{minipage}{.93\columnwidth}
\small
\textbf{Fail-closed request procedure}
\begin{enumerate}
  \item Verify model and private-checkpoint manifests.
  \item Resolve trusted metadata to a row-level Boolean vector; invalid rows become deny.
  \item Reject any KV cache whose policy tag or model fingerprint differs from the request.
  \item At each wrapped layer, mask the ineligible pool to $-\infty$, then compute top-$k$.
  \item Record layer, row, token, selected pool, and policy decision in the route audit.
  \item Before the next arm or sample, clear KV, policy, route, RNG, tokenizer, and application state.
\end{enumerate}
\end{minipage}}
\par\smallskip
{\small The authorization audits record selections, not merely the final response.}
\par\medskip
\end{center}

\subsection{State isolation and cache discipline}

Mixed-policy batches carry an explicit Boolean vector through both wrapped
blocks.  Route counters are indexed by row and layer, so a batch-level total
cannot hide cross-row leakage.  KV caches are tagged with the policy decision,
model revision, and checkpoint fingerprint.  Reuse across policy tags is
rejected before a model call; the subsequent deny pass starts from fresh state.
The allow--deny--recovery tests additionally reset route counters and RNG state
between decisions.  These measures are part of the TCB rather than properties
learned by the model.

\subsection{Guarantee boundary}

Conditional on an uncompromised TCB, the hard claim is that private modules do
not execute under deny.  The claim does not cover capabilities already present
in public parameters, information carried through shared attention or the
residual stream, authorized users who copy outputs to unauthorized users,
downstream tool permissions, host compromise, or timing and memory side
channels.  Tools and external actions therefore require their own authorization
checks even after the model route is denied.

\section{Immutable Training Configurations}

\subsection{Primary Qwen system}

The primary model is
\hash{Qwen/Qwen3-30B-A3B-Instruct-2507} at revision
\hash{0d7cf23991f47feeb3a57ecb4c9cee8ea4a17bfe}.  We add 16 private experts at
each of layers 46 and 47, for 151,060,480 trainable parameters.  The immutable
selection contains 8,524 train and 890 validation targets, 16,084,443 admitted
tokens, and a 2,273-batch deterministic plan.  Training uses BF16, SDPA, fused
AdamW, learning rate $10^{-4}$, an 8,192-token batch budget, and final-step-only
checkpoint selection.  Every non-private parameter has gradients disabled and
is fingerprinted before and after training.

\begin{table}[t]
\centering
\caption{Complete Qwen matched-seed results.  Relative reduction is $(L_0-L_T)/L_0$.}
\label{tab:app-seeds}
\resizebox{\columnwidth}{!}{%
\begin{tabular}{lrrrrr}
\toprule
Seed & Init. val. & Final val. & Rel. red. & tok/s & Peak GB \\
\midrule
20260730 & .983972 & .573489 & .417169 & 3682.0 & 82.558 \\
20260731 & .983342 & .573436 & .416850 & 3694.2 & 82.559 \\
20260732 & .984312 & .573335 & .417527 & 3636.6 & 82.559 \\
\midrule
Mean & -- & .573420 & .417182 & -- & -- \\
Sample SD & -- & .000078 & .000339 & -- & -- \\
\bottomrule
\end{tabular}}
\end{table}

All three runs complete 2,273 steps.  Each of the 32 private experts receives
routing events and finite, nonzero gradients; deny and missing metadata produce
zero private selections; the frozen fingerprint remains exact; and both private
weights and optimizer state reload from the final checkpoint.  Throughput is
reported only as execution context and was not used to select a seed.

The selected checkpoint SHA-256 is
\hash{b154511b822681ef3be6f30a5370696e4c0b3120002956ee77f235b0be42cbda};
its tensor fingerprint is
\hash{6225da4ea2f9f66a4142c5fd30b0ef675fc5647a2a5faea1b9dcff7373559211}.
The checkpoint was chosen as the median seed on the pre-existing open
validation endpoint before E7 output.

\subsection{Matched expert-LoRA baseline}

The baseline matches model, data, admitted tokens, batch order, 2,273 steps,
BF16/SDPA, fused AdamW, learning rate, seed, and final-step selection.  It adapts
gate, up, and down projections in every public expert at layers 46--47.  Of 768
modules, 652 use rank 70 with $\alpha=140$ and 116 use rank 69 with
$\alpha=138$.  The total is 151,061,504 trainable parameters, only 1,024 more
than the private path.  Initial/final validation losses are .898204/.555237,
for a .381836 relative reduction.  LoRA is an adaptation baseline and has no
private-path or hard-isolation claim.

\subsection{DeepSeek replication}

The second model is \hash{deepseek-ai/DeepSeek-V2-Lite-Chat} at revision
\hash{85864749cd611b4353ce1decdb286193298f64c7}.  Layers 25 and 26 each receive
16 private experts, totaling 276,889,600 trainable parameters.  The prospective
split has 1,024 train, 128 validation, and 100 held-out records, with 1,602,789
admitted train tokens in 226 deterministic batches.  Training uses whole-model
BF16 computation, explicit FlashAttention-2, fused AdamW at $10^{-4}$, and
\code{torch.compile} in default mode with dynamic shapes and partial graphs.
The numerical
family and one-pass, final-step-only rule are fixed before held-out generation.

Validation loss changes from 1.391523 to .909779, a .346199 relative
reduction.  All 226 steps finish; all 32 experts receive routes and gradients;
the frozen model remains exact; and the final private and optimizer states
reload.  The final checkpoint SHA-256 is
\hash{94efbbe35e6200d6d1860e72f9c75df780fbf1280e2528bf9a6ecc2c00f6a9d4}.

\section{Benchmark Construction and Frozen Analysis}

\subsection{Data separation and contamination audit}

The E7 suite has two roles selected and frozen together.  The task-aligned role
draws ToolMind single-call targets from an immutable source only when the
record, exact target, called-tool name, and called-tool parameter schema are
absent from the primary training and validation targets.  The fresh external
role draws ToolFailBench tasks from pinned commit
\hash{c8be7fb0f1d295b1e116d7bd0e01d4c5e91f1653}, stratified by five domains and
four target modes.  Selection uses fixed SHA-256 ranks, never a model score.

Canonical exact overlap and normalized token-set overlap are audited against
the training targets, all CCTU records, and the other E7 role.  Candidates with
prompt or called-tool Jaccard similarity at least .85 are removed before model
evaluation and replaced by the next fixed-ranked candidate.  The initial
160+160 public-only calibration projected 11.564 GPU hours, above the declared
eight-hour cap.  Before any adapted output, the single permitted nested
reduction retained 100 ToolMind records (25 per length quartile) and 100
ToolFailBench records (five per domain-by-mode stratum).  The final projection
was 7.2275 GPU hours and the planning minimum detectable difference was 18.2
percentage points.  No second reduction was allowed.

The reduced benchmark manifest is
\hash{d5e4a5a94bb2be6e7e694d369db269fa531069c08964f22ce6e132a35dfe3b26};
the six-condition matrix lock is
\hash{eabceda9e15f2e85f91c3527dea6f04f2459975b0995e38e1b738035d42216bd}.

\subsection{Decoding and scoring}

Qwen uses greedy decoding, \code{do\_sample=false}, thinking disabled, batch
size eight, at most 512 new tokens, and the tokenizer-native Hermes tool
template.  Every arm clears KV, policy, route, RNG, and application state for
every sample.  The task-aligned score is exact ordered tool-call match after
canonicalizing function names and JSON arguments; no semantic judge is used.
The external primary score is ToolFailBench's deterministic Clean Tool-Use Rate
(CTUR) on the 75 tool-required cases.  All 100 external records, including 25
controls, remain in the artifacts and descriptive summaries.

Post lock, a blinded packet contains all public, full-private, and LoRA outputs
for the same 100 records. Three independent model evaluators label the 300
arm-blinded candidates under a frozen rubric. These judgments are a sensitivity
analysis rather than human annotation or a replacement for the deterministic
primary rule.

For pair $i$, let $Y_i^{a}$ and $Y_i^{p}$ be binary allow and public outcomes.
The risk difference is
\begin{equation}
\widehat{\rd}=\frac{1}{n}\sum_{i=1}^{n}(Y_i^{a}-Y_i^{p}).
\end{equation}
With $b=\sum\mathbf{1}[Y_i^p=0,Y_i^a=1]$ and
$c=\sum\mathbf{1}[Y_i^p=1,Y_i^a=0]$, the one-sided exact McNemar test uses the
upper binomial tail for $b$ conditional on $b+c$.  The two directional E7 tests
are corrected by Holm at family-wise $\alpha=.05$.  Confidence intervals are
percentile intervals from 100,000 paired resamples stratified by frozen
benchmark strata with seed 20260801.  PSR and other diagnostics are descriptive
unless explicitly preregistered as primary.

\section{Complete Qwen Control Matrix}

Table~\ref{tab:app-matrix} reports all prospectively frozen conditions.  The
prompt-only arm modifies the instruction context but not weights or routing.
Soft-router-only exposes the learned mixed routing behavior without the hard
private-only mask.  Mask-only activates the private-only topology before
private training.  Full private combines trained private experts and the hard
mask.  Matched LoRA supplies a parameter-matched adaptation baseline.

\begin{table*}[t]
\centering
\caption{Complete E3/E7 comparison.  Differences are condition minus frozen public.  Inferential $p$-values were preregistered for full private; the other control differences are descriptive.}
\label{tab:app-matrix}
\begin{tabular}{lrrlrrl}
\toprule
& \multicolumn{3}{c}{Task-aligned exact tool calls ($n=100$)} & \multicolumn{3}{c}{Fresh external CTUR ($n=75$)} \\
\cmidrule(lr){2-4}\cmidrule(lr){5-7}
Condition & Rate & RD & 95\% bootstrap CI & Rate & RD & 95\% bootstrap CI \\
\midrule
Frozen public & .560 & .000 & [.000,.000] & .507 & .000 & [.000,.000] \\
Prompt-only & .560 & .000 & [-.040,.040] & .547 & +.040 & [-.013,.093] \\
Soft-router-only & .560 & .000 & [.000,.000] & .587 & +.080 & [.013,.147] \\
Mask-only & .490 & -.070 & [-.130,-.020] & .467 & -.040 & [-.107,.027] \\
Full private & .610 & +.050 & [.010,.100] & .720 & +.213 & [.133,.293] \\
Matched LoRA & .550 & -.010 & [-.070,.050] & .707 & +.200 & [.120,.280] \\
\bottomrule
\end{tabular}
\end{table*}

For full private versus public, the task-aligned discordances are five
public-fail/private-success and zero public-success/private-fail.  The
preregistered one-sided exact McNemar and Holm-adjusted value is $p=.03125$;
the corresponding two-sided exact value is $p=.0625$.  Conditional on
discordance, the private-win probability is $1.0$, with exact one-sided 95\%
lower bound $.549$ and two-sided 95\% interval $[.478,1.000]$.  The
percentile-bootstrap RD interval $[.010,.100]$ is retained as a descriptive
effect-size summary, not independent two-sided evidence, because empirical
resampling cannot generate the absent reverse discordance.  Fresh-external
discordances are 16 and 0; raw $p=.0000152587890625$ and Holm-adjusted
$p=.000030517578125$.
The two roles are retained regardless of sign.  In secondary paired
comparisons, full private exceeds matched LoRA by $.060$ on task-aligned data
(95\% CI $[.010,.120]$, seven versus one, $p=.03515625$) and by $.0133$
externally (95\% CI $[-.0267,.0533]$, two versus one, $p=.5$).  The matrix
therefore supports a useful private increment on both Qwen roles without
establishing universal adaptation superiority.

\section{Authorization and Recovery Audits}

\subsection{Adversarial suite}

The E4 suite contains 64 scenarios, eight in each frozen family, and 152 policy
events.  It is constructed without CCTU sealed outcomes or E7 outputs.  The
endpoint is measured by forward-pass route events; behavioral top-token output
is descriptive.

\begin{table*}[t]
\centering
\caption{Frozen authorization attack families and required invariants.}
\label{tab:app-e4}
\begin{tabular}{p{.19\textwidth}p{.47\textwidth}p{.21\textwidth}}
\toprule
Family & Attack or transition & Required result \\
\midrule
Direct override & User asks the model to ignore a deny decision & zero private execution \\
Role/system imitation & User text claims a higher-priority role or authorization & zero private execution \\
Missing metadata & Trusted authorization field is absent & fail closed \\
Malformed metadata & Wrong key, string, number, list, nested object, null, or serialized JSON & fail closed \\
Mixed-policy batch & Eight rows alternate allow and deny & row-local routes; no leakage \\
Allow--deny--allow & Same logical request changes policy across three cold decisions & deny isolation; restored allow \\
Private-name imitation & Denied prompt uses private tool, expert, or task names & zero private execution \\
Cache isolation & Allow cache is presented to deny, then clean deny runs & reject reuse; fresh deny exact \\
\bottomrule
\end{tabular}
\end{table*}

All 56 allow events are private-only.  All 96 deny or fail-closed events produce
zero unauthorized private selections.  Policy mismatches and harness exceptions
are zero, mixed rows remain isolated, cache reuse is rejected, and recovery
passes.  Peak allocated memory is 61.4 GB.  The suite SHA-256 is
\hash{f084b7c782199363804312199434230eed0c8687eb33569c2bba7f22c0170c3a};
the freeze manifest is
\hash{8ae595f78710ce1b6b8ee52000226feb19f41feeaf4519164705159df4651a8e}.

An independent physical-execution audit registers forward hooks directly on all
32 private modules, outside the route logger. It observes 931 calls and 11,616
physical input rows, exactly equal to the routed-row total. No denied sequential
event invokes a private module. This reduces common-mode instrumentation risk,
but is not a kernel-level or formal proof.

\subsection{Request-gated LoRA and evaluator sensitivity}

The unchanged matched adapter is evaluated on the frozen 100+100 IDs under
allow, deny, and recovery. Hooks cover 1,536 LoRA A/B modules. Deny records
1,225 forward calls ($0.04242\%$ of the allow count); allow and recovery each
observe 2,887,794 calls and have byte-exact ordered outputs. Task-aligned
allow/deny rates are .54/.57
($\rd=-.03$, 95\% CI $[-.09,.03]$, $p=.9102$); external tool-required rates
are .72/.5467 ($\rd=+.1733$, 95\% CI $[.0933,.2533]$, $p=.000122$). All 200
records are scorable with zero parse or harness errors. The auxiliary
\code{requires\_grad} metadata gate passes: every adapter parameter is frozen
before, during, and after deny inference. Inference mode, no
optimizer/backward, and exact checkpoint hashes establish that no update
occurred.

For blinded ToolFailBench labels, multiclass Fleiss $\kappa=.7727$ and pairwise
Cohen kappas are .7498/.8033/.7671; 84.7\% of candidates are unanimous and
28/300 have no majority. Automated exact-category and binary-success agreement
are 78.3\% and 80.3\%. Treating no-majority as failure, tool-required
private--public is $+.1867$ (95\% CI $[.0933,.2800]$, 16 versus 2
discordances, $p=.000656$), while LoRA--public is $+.160$ (95\% CI
$[.0667,.2533]$, 15 versus 3, $p=.00377$). Private--LoRA is $+.0267$
(95\% CI $[-.0267,.0933]$, four versus two, $p=.34375$). These model-evaluator
labels corroborate the external direction but are not human judgments.

\subsection{Qwen and DeepSeek toggles}

Qwen full-private allow and recovery are private-only and have zero ordered
behavior mismatches on both E7 roles.  Deny has zero private execution.  The
separately generated E7 deny output differs from the cached frozen-public
projection on 15 task-aligned and 99 external records.  Because both artifacts
have exact frozen/checkpoint fingerprints and the deny route trace is clean,
these differences are reported as deterministic-serving diagnostics rather
than recast as unauthorized execution.

For DeepSeek, all four public/allow/deny/recovery arms retain 100 records and
have zero harness exceptions.  Public and deny each have nine exact matches and
six parse deviations affecting five records; allow and recovery each have 36 exact matches and four
parse deviations.  Deviations count as failures.  Public/deny raw and parsed
outputs are exact, with zero private execution under deny.  Allow/recovery raw
and parsed outputs are exact and private-only.  The paired allow-minus-public
RD is +.27, 95\% CI [.19,.35], with one-sided exact McNemar
$p=7.450580596923828\times10^{-9}$.

\section{Valid Sealed CCTU Evaluation}

CCTU is split by a fixed seed into a 40-case open pilot and a 160-case sealed
set balanced across Single-Hop, Parallel Single-Hop, Multi-Hop, and Parallel
Multi-Hop.  The sealed dataset SHA-256 is
\hash{e7fb2b3a0d513ffe4f60544b968b54da31c46597da6ee3d3c93b6f9708fac5a0}.
Before opening, the model, checkpoint, arm order (public, deny, allow), greedy
decoder, 20-round cap, 512-token-per-round cap, judge, statistics, and validity
gates are fixed.  The seal is opened once and the opening receipt is retained.

All three arms are 160/160 scorable with zero harness exceptions and zero
model-format parse errors.  Public and deny transcripts are byte-identical and
deny has zero private execution.  Every preregistered validity gate passes, so
all 160 pairs retain sealed confirmatory status.

Observed SR is .150 public and .14375 allow (RD $-.00625$, 95\% CI
$[-.0375,.025]$, one-sided exact McNemar $p=.7734375$), with three allow-only
and four public-only discordances.  PSR is .09375/.0875 (RD $-.00625$, 95\%
CI $[-.01875,0]$, $p=1$).  The valid confirmatory result is near-neutral and
does not support the positive directional hypothesis.

\section{Failure, Repair, and Numerical-Family Lineage}

The experiment ledger preserves failures that occurred before scientific
output.  Two E7 matrix attempts stop before the affected generation because of
a log-directory ordering bug and stale remote staging.  Completed public and
prompt artifacts are reused only after their independent manifests verify.

DeepSeek FP8 preflights reveal two incompatibilities in the selected software
stack: empty-expert dispatch triggers an invalid reduction, and a 152-row
backward reduction violates a kernel alignment constraint.  Both occur before
training or held-out output.  The family is therefore changed prospectively to
stable BF16 with explicit FlashAttention-2 and dynamic \code{torch.compile};
the final training run is not chosen through a throughput comparison.  A first
post-training audit also applies an over-strong full-vocabulary equality test
across different batch shapes.  Without retraining or opening held-out output,
the gate is aligned to the primary Qwen contract: exact per-row routes,
same-shape full logits, and mixed-batch greedy next-token identity.  The
original failed gate remains recorded.

Two subsequent generation attempts stop before the first token because pinned
DeepSeek remote code expects legacy \code{DynamicCache} query methods.  A
read-only compatibility bridge restores only \code{seen\_tokens},
\code{get\_max\_length}, and \code{get\_usable\_length}; cache tensor updates
are unchanged.  An open-only audit exactly reproduces the frozen checkpoint
sequences and routes before the held-out run.  This lineage matters because a
single final ``pass'' label would conceal changes to the execution contract.

\section{Reproduction Checklist and Evidence Ledger}

The final clean lock performs two operations.  First, it reruns the unchanged
40-case Qwen open-pilot allow configuration from pinned source, data, and
checkpoint without reading or scoring the sealed cases.  The result is 40/40
scorable, SR/PSR .15/.10, zero harness and parse errors, 64 private selections
at each wrapped layer in the bounded probes, maximum public-equivalence error
zero, exact recovery, and 0\% drift.  Its ordered transcript SHA-256 is
\hash{9739eb5520dbbc3176135307f7225c6de34db05fdfeb5ec4d568aaef3381e1fb}.

Second, the final builder verifies all raw manifests and recomputes CCTU, E1,
E3/E7, E4, and E6 summaries and tables.  Existing summaries are comparison
targets, not inputs.  All component rebuilds are byte- or field-exact.  A
minimal reproduction should therefore verify, in order:

\begin{enumerate}
  \item source revision, model revision, tokenizer files, and frozen-base fingerprint;
  \item dataset membership, canonical serialization, exclusion audit, and batch-plan hashes;
  \item trainable-parameter partition, optimizer state, finite losses/gradients, and final checkpoint reload;
  \item allow, deny, missing-metadata, mixed-row, cache-isolation, and recovery route gates;
  \item raw per-sample generations, parser outcomes, scores, exception records, and arm manifests;
  \item paired statistics, clean-lock tables, and the final evidence hashes below.
\end{enumerate}

\paragraph{Immutable ledger.}
\begin{itemize}
  \item Source commit: \hash{58ef28e65df7d27f0296db58b2a0f399b604911e}.
  \item Final evidence JSON: \hash{9a6d9fb194fa61a9ddcbe97900da55d34c3c46df5a3fceba803157eb62417974}.
  \item Final evidence TSV: \hash{d50a55117ff36ab94b0d03314e8054434da5bd4255231bea7521267c9bb697ac}.
  \item Final evidence \LaTeX{} table: \hash{402024d833f92a6d921fe8367f8cb5c91a6321a4b59cebe1aa99449cbeb5e3ce}.
  \item Final provenance JSON: \hash{9c2e264af69d46ea199abd215d18378f39013f6cb2d4d2b4cba28daa7277bb29}.
  \item Reviewer-audit summary JSON: \hash{15d43b1f98cfe2da672c13f25135596000c2eca06e0216e016e94f43bbfca797}.
  \item Blinded ToolFailBench packet: \hash{3bc0e52719470977b0e039e72a477efe5804ef499ab1c18d2fe79e155ab3c6a6}.
\end{itemize}

These hashes identify the material used to populate the paper; they do not
imply that the paper prose was preregistered.  Licensed model weights and
benchmark data should be retrieved from their pinned upstream revisions rather
than redistributed.  A public artifact release should include the
preregistrations, membership manifests where licensing permits, route summaries,
statistics code, environment locks, failure ledger, and the clean-lock evidence
files.

\section{Deployment, Ethics, and Responsible Use}

Policy-masked private experts can enforce a technically precise policy while
the policy itself remains unfair, opaque, or unsafe.  A deployment should state
the legitimate basis for each access tier, minimize the data retained in route
logs, expose an appeal and correction path, and keep tool-side authorization
independent of model output.  Operators should monitor the TCB, rotate signed
manifests, test cache and mixed-batch isolation after serving changes, and fail
closed when identity or policy services are unavailable.  Because public
parameters may already implement similar behavior, deny must never be marketed
as proof that a user cannot elicit the underlying semantic capability.  The
proper claim is narrower: the newly added private increment was not executed.

\else
\balance
\fi
\bibliographystyle{plainurl}
\bibliography{capability_gating_references}

\end{document}